\documentclass[twocolumn]{autart}    
\usepackage{array}
\newcolumntype{P}[1]{>{\centering\arraybackslash}p{#1}}
\usepackage{algorithm}
\usepackage{algorithmicx}
\usepackage{algpseudocode}
\usepackage{tabu}
\usepackage{amssymb}
\usepackage{mathrsfs}
\usepackage{amsmath,cite}
\usepackage{chemarrow}
\usepackage{hyperref}
\usepackage{url}
\usepackage{color}
\usepackage{graphicx}          
\usepackage{multirow}
\usepackage{graphicx}
\usepackage{epstopdf}

\begin{document}
	
	\begin{frontmatter}
		
		\title{High Precision Variational Bayesian Inference of Sparse Linear Networks \thanksref{footnoteinfo}} 
		\thanks[footnoteinfo]{For correspondence:  yye@hust.edu.cn.}
		
		\author[LUX]{Junyang Jin},   
		\author[WU]{Ye Yuan}$^{,\star}$,        
		\and \author[LUX]{Jorge Gon\c{c}alves}
		
		\address[LUX]{The Luxembourg Centre for Systems Biomedicine, avenue du Swing, 4367 Belvaux, Luxembourg}
		\address[WU]{School of Automation, Huazhong University of Science and Technology, 430074, Hubei, China}        


		\begin{abstract}		
			Sparse networks can be found in a wide range of applications, such as biological and communication networks. Inference of such networks from data has been receiving considerable attention lately, mainly driven by the need to understand and control internal working mechanisms. However, while most available methods have been successful at predicting many correct links, they also tend to infer many incorrect links. Precision is the ratio between the number of correctly inferred links and all inferred links, and should ideally be close to $100\%$. For example, $50\%$ precision means that half of inferred links are incorrect, and there is only a $50\%$ chance of picking a correct one. In contrast, this paper infers links of discrete-time linear networks with very high precision, based on variational Bayesian inference and Gaussian processes. Our method can handle limited datasets, does not require full-state measurements and effectively promotes both system stability and network sparsity. 
On several of examples, Monte Carlo simulations illustrate that our method consistently has $100\%$ or nearly $100\%$ precision, even in the presence of noise and hidden nodes, outperforming several state-of-the-art methods. 
The method should be applicable to a wide range of network inference contexts, including biological networks and power systems.

		\end{abstract}
		
	\end{frontmatter}
	
	\section{INTRODUCTION}\label{sec:introduction}
	
	In systems biology, mathematical modelling has been central to the study of biological networks. Dynamical models are frequently developed to predict the behaviour of systems in response to external or internal stimulus for example, drug treatment or mutation. Yet only input-output dynamics are learned without exploring the topology, whereas in many other applications, comprehensive knowledge of the network topology is required. For example, the information about the structure of control systems is essential for fault diagnosis. Hence, both the inference of system dynamics and the detection of network topology are important.
	
	Precision is the ratio between the number of correctly inferred links and all inferred links. It indicates whether inferred networks can be trusted. For example, if precision is close to or even below $50\%$, it is impossible to tell which inferred links are correct. Therefore, precision should be the first priority when solving network inference. However, most state-of-the-art methods can rarely achieve $100\%$ precision, meaning that not all inferred links are correct. The motivation of this work is to develop a method that prioritises precision over true positive rate (TPR). Several examples consistently achieved $100\%$ or nearly $100\%$ precision outperforming other state-of-the-art methods.
	
	Sparsity and stability are fundamental properties of most real-world networks. Communication networks, as artificial systems, are designed to be stable for robust operation and sparse to reduce energy consumption. Thus, sparsity and stability are primary constraints in network inference, especially in the case of limited data source or high amount of noise. When dealing with practical networks, often, not all the nodes in the network can be measured, because of either high experimental cost or technical limitations. The difficulty here is that many inference methods nonetheless assume full-state measurements. It is important to reconsider this issue carefully.
	
	In recent years, kernel-based methods have become popular in the system identification community~\cite{nonp1}. For linear systems, the methods effectively impose system stability and greatly simplify the estimation of model complexity. Kernel-based methods have successfully identified SISO continuous linear time invariant (LTI) models~\cite{nonp1} and been further extended to discrete LTI systems\cite{nonp1,nonp3, nonp4}. In particular, kernel-based methods have been used to infer sparse networks described by Granger causality~\cite{nonp}. They have been further developed to infer the so-called sparse plus low rank networks where it is assumed that the majority of nodes can be described by a few other components that are not accessible for observation~\cite{sl}. In addition, system identification of a variety of model classes have been considered: the models include NFI, NARX, NARMAX, linear parameter-varying (LPV) Box-Jenkins models, Hammerstein models, and cascaded linear systems~\cite{nonp9, boxj, uncer, uncer1}. System dynamics and network topology are controlled by the hyperparameters of kernel functions. Under the Bayesian paradigms, kernel-based methods apply emprical Bayes to estimate hyperparameters (KEB). By incorporating Automatic Relevance Determination (ARD), kernel-based methods are able to enforce sparsity, where the sparsity profile of the solution implies network topology~\cite{nonp, nonp3,sl}. Nevertheless, this framework is not ideal for topology detection. Due to local optimal solutions, it is very difficult to achieve $100\%$ precision.
	

	Variational inference (VI), as empirical Bayes, is a Bayesian deterministic approximation technique that has been applied to a number of cases, including sparse regression models~\cite{pattern, VI6} and neural networks~\cite{VI3, VI4, VI7}. Instead of estimating hyperparameters directly, VI searches for an approximation of the posterior distribution of hyperparameters (functional estimation). With well-posed models of exponential families, VI is as efficient computationally as empirical Bayes~\cite{VI2}. Whilst rigorous evaluation remains elusive,   Monte Carlo simulations show that VI can be more accurate than empirical Bayes~\cite{VI1, VI2}. More importantly, VI is able to estimate model evidence for each possible model structure: this enables evaluation of the relative confidence between models. However, VI is barely used in kernel-based system identification, probably due to nonlinearities introduced by kernel functions (non-gaussianity). VI does no hold a closed-form update in the algorithm, and it has to deal with high-dimensional ill-conditioned covariance matrices, which seriously increases computational burden and degrades numerical stability. Nevertheless, thanks to recent developments on analysis of kernel functions~\cite{nonp8}, VI achieves considerably higher efficiency and robustness by using special kernel functions (Tuned/Correlated kernel).
	
	This paper combines Gaussian processes and VI to infer sparse networks. Dynamical structure functions are used to describe sparse linear networks where the information of hidden nodes is encoded via transfer functions. By expressing DSF models in a non-parametric way, the system can be identified without knowing the number of hidden nodes and the connectivity among them. VI is employed to identify system dynamics and infer network topology. Moreover, by applying backward selection strategies, the proposed method encourages high inference precision. Monte Carlo simulations show that our method produces more reliable networks than KEB under various experimental conditions, such as different topologies, noise levels, kernel functions and number of data points. Most importantly, the proposed method always achieves $100\%$ or nearly $100\%$ precision so that almost all inferred links are correct.

	The paper is organized as follows. Section~\ref{sec:Bayesian} introduces variational inference algorithms. Section \ref{sec:Model} introduces dynamical structure function and formulates the full Bayesian model. Section~\ref{sec:inference} discusses network inference using variational inference and analyses algorithm properties. Section~\ref{sec:Simulation} compares the method with other approaches via Monte Carlo simulations. Finally, Section~\ref{sec:Conclusion} concludes and discusses further developments in this field.
	
	\emph{Notation}: The notation in this paper is as follows. $I_n$ denotes a $n\times n$ identity matrix. For $L\in{R}^{n\times n}$, $diag\{L\}$ denotes a vector which consists of the diagonal elements of matrix $L$. $[L]_{ij}$ presents the $ij$th entry and $\overline{L_i^j}$ denotes the $i$th $j\times j$ diagonal block of $L$. For a series of matrices, $\{L_i| i=1,...,n\}$, $blkdiag\{L_1,...,L_n\}$ denotes a block diagonal matrix. For $l\in{R}^{n}$, $diag\{l\}$ denotes a diagonal matrix whose diagonal elements come from vector $l$. $l_{ij}$ denotes the $j$th element of the $i$th group of $l$. A vector, $y(t_1 : t_2)$ denotes a row vector $\left[\begin{array}{cccc}y(t_1)&y(t_1 +1)&\cdots&y(t_2)\end{array}\right]$. $\mathcal{N}(x|m,\Sigma)$ denotes a Gaussian distribution of $x$ with mean $m$ and covariance matrix $\Sigma$. $asc\{a_1,...,a_n\}$ means to rearrange the elements in an ascending order of the magnitude.
	
	\section{OVERVIEW OF VARIATIONAL INFERENCE}\label{sec:Bayesian}
	
	Variational inference (VI) approximates a full Bayesian model analytically so that the intractable marginalization or expectation can be easily calculated~\cite{mur, VI5}. Empirical Bayes was more frequently used in the kernel-based system identification, where system dynamics are the main concern. Under the context of network inference, model selection (detection of network topology) is another important aspect. Model evidence is usually required to compare different model structures. Whilst empirical Bayes does not evaluate model evidence, variational inference generates a lower bound for it, which motivates advanced strategies for model selection. 
	Consider a model structure $\mathcal{M}_k$ with model parameters, $\theta$ and data, $y$. The model evidence, $p(y|\mathcal{M}_k)$ is expressed as:
	\begin{equation}
	\begin{aligned}
	p(y|\mathcal{M}_k)=\int p(y|\theta,\mathcal{M}_k)p(\theta|\mathcal{M}_k)d\theta.\\
	\end{aligned}
	\end{equation}    
	Assuming an arbitrary distribution $Q(\theta|\mathcal{M}_k)$ is used to approximate $p(\theta|y,\mathcal{M}_k)$, we have:
	\begin{equation}
	\begin{aligned}
	&\ln p(y|\mathcal{M}_k)=\ln \frac{p(\theta|y,\mathcal{M}_k)p(y|\mathcal{M}_k)}{p(\theta|y,\mathcal{M}_k)}\\
	&=\ln \frac{p(y|\theta,\mathcal{M}_k)p(\theta|\mathcal{M}_k)}{p(\theta|y,\mathcal{M}_k)}\\
	&=\int Q(\theta|\mathcal{M}_k)\ln \frac{p(y|\theta,\mathcal{M}_k)p(\theta|\mathcal{M}_k)}{p(\theta|y,\mathcal{M}_k)}d\theta\\
	&=\int Q(\theta|\mathcal{M}_k)\ln \frac{p(y|\theta,\mathcal{M}_k)p(\theta|\mathcal{M}_k)}{Q(\theta|\mathcal{M}_k)}d\theta\\
	&+\int Q(\theta|\mathcal{M}_k)\ln \frac{Q(\theta|\mathcal{M}_k)}{p(\theta|y,\mathcal{M}_k)}d\theta\\
	&=L[Q(\theta|y,\mathcal{M}_k)]+KL[Q(\theta|\mathcal{M}_k)||p(\theta|y,\mathcal{M}_k)].
	\end{aligned}
	\label{BF}
	\end{equation}  
	where
	\begin{equation}{\tiny
	\begin{aligned}
	&L[Q(\theta|y,\mathcal{M}_k)] =\int Q(\theta|\mathcal{M}_k)\ln \frac{p(y|\theta,\mathcal{M}_k)p(\theta|\mathcal{M}_k)}{Q(\theta|\mathcal{M}_k)}d\theta\\
	&KL[Q(\theta|\mathcal{M}_k)||p(\theta|y,\mathcal{M}_k)]=\int Q(\theta|\mathcal{M}_k)\ln \frac{Q(\theta|\mathcal{M}_k)}{p(\theta|y,\mathcal{M}_k)}d\theta.
	\end{aligned}}
	\end{equation}        
	The second term of \eqref{BF} is the Kullback-Leibler (KL) divergence between $Q(\theta|\mathcal{M}_k)$ and $p(\theta|y,\mathcal{M}_k)$.  KL divergence is non-negative and equal to zero if and only if $Q(\theta|\mathcal{M}_k)=p(\theta|y,\mathcal{M}_k)$. Therefore, KL divergence measures the difference between the true and the approximate distributions. More importantly, since $KL[Q(\theta|\mathcal{M}_k)||p(\theta|y,\mathcal{M}_k)]\geq 0$, $\ln p(y|\mathcal{M}_k)\geq L[Q(\theta|y,\mathcal{M}_k)]$, meaning $L[Q(\theta|y,\mathcal{M}_k)]$ is the lower bound of the logarithm of model evidence. Hence, it can be used as an approximation of model evidence for model selection.
	
	VI employs KL divergence as the metric to measure the accuracy of the approximation. Therefore, the goal is to find the optimal $Q(\theta|y,\mathcal{M}_k)$ that minimizes the KL divergence:
	\begin{equation}
	\begin{aligned}
	Q_{opt}(\theta|y,\mathcal{M}_k) = \arg\min_{Q}KL[Q(\theta|\mathcal{M}_k)||p(\theta|y,\mathcal{M}_k)].\\
	\end{aligned}
	\end{equation}   
	Equation \eqref{BF} implies that  $KL[Q(\theta|y,\mathcal{M}_k)||p(\theta|y,\mathcal{M}_k)]=\ln p(y|\mathcal{M}_k)-L[Q(\theta|y,\mathcal{M}_k)]$ where $\ln p(y|\mathcal{M}_k)$ is independent on $Q$. Hence, the optimization problem is equivalent to maximizing the lower bound $L[Q(\theta|y,\mathcal{M}_k)]$:
	\begin{equation}
	\begin{aligned}
	Q_{opt}(\theta|y,\mathcal{M}_k) = \arg\max_{Q} L[Q(\theta|y,\mathcal{M}_k)].\\
	\end{aligned}
	\label{VI_opt1}
	\end{equation}   
	Without further constraints on $Q(\theta|y,\mathcal{M}_k)$, the solution is $Q(\theta|y,\mathcal{M}_k)=p(\theta|y,\mathcal{M}_k)$, which offers no help to resolve intractable Bayesian estimation. To relax the complicated Bayesian model, $Q(\theta|y,\mathcal{M}_k)$ is assigned with a simple structure. VI expresses $Q(\theta|y,\mathcal{M}_k)$ in a factorized form based on the mean field theorem in physics:
	\begin{equation}
	\begin{aligned}
	Q(\theta|y,\mathcal{M}_k) = \prod q(\theta_i|\mathcal{M}_k).\\
	\end{aligned}
	\end{equation}     
	where $q(\theta_i|\mathcal{M}_k)$ are independent distributions for each element of $\theta$. Consequently, problem~\eqref{VI_opt1} becomes:
	\begin{equation}
	\begin{aligned}
	&Q_{opt}(\theta|y,\mathcal{M}_k) = \arg\max_{Q} L[Q(\theta|y,\mathcal{M}_k)]\\
	&\text{subject\ to:}~Q(\theta|y,\mathcal{M}_k) = \prod q(\theta_i|\mathcal{M}_k)\\
	&\int q(\theta_i|\mathcal{M}_k)d\theta_i=1,\ i=1, 2, ...\\
	\end{aligned}
	\label{op}
	\end{equation}     
	By substituting the constraints, the cost function becomes convex with respect to each factor, $q(\theta_i|\mathcal{M}_k)$. According to the theory of variational calculus, the solution to the problem is:
	\begin{equation}
	\begin{aligned}
	\ln q(\theta_i|\mathcal{M}_k) = E_{j\neq i}[\ln p(y,\theta)]+c_{\theta_{j\neq i}}.\\
	\end{aligned}
	\label{q}
	\end{equation}  
	where the expectation is conducted with respect to the factors, $q(\theta_j|\mathcal{M}_k)$ ($j\neq i$). $c_{\theta_{j\neq i}}$ is a term independent on $\theta_i$. Although equations in \eqref{q} indicates the consistency conditions of the optimal solution to problem \eqref{op}, they cannot be solved analytically. To seek for the solution, the factors are updated cyclically following the scheme of the coordinate descent method. Since the cost function is convex, convergence is guaranteed~\cite{pattern}.

	\section{MODEL FORMULATION}\label{sec:Model}
	\subsection{The dynamical structure function}
	The sparse network of $n$ nodes is described by a linear state space model as follows:
	\begin{equation}
	\begin{aligned}
	x(t+1) &= Ax(t)+B_uu(t)+B_ee(t)\\
	y(t) &= Cx(t).
	\label{ssm}
	\end{aligned}
	\end{equation}
	where $x\in \mathcal{R}^n$ are states of the system, each of which represents a node. $u\in \mathcal{R}^m$ denote inputs. $y\in \mathcal{R}^p$ present the measurements of the states. $e\in \mathcal{R}^q$ are i.i.d white Gaussian noise with zero mean and covariance matrix $P_e$. Without loss of generality, $P_e$ is assumed to be diagonal. If the covariance matrix is full, one can decompose the matrix using singular value decomposition (SVD) as $P_e=R\Sigma R'$. By replacing $B_e$ with $B_eR$, noise $e$ have a diagonal covariance matrix. $A\in \mathcal{R}^{n\times n}$, $B_u\in \mathcal{R}^{n\times m}$, $B_e\in \mathcal{R}^{n\times q}$ and $C\in \mathcal{R}^{n\times p}$ are system matrices.
	
	It is normal in practice that some of the nodes are unobservable (hidden states). For example, in a gene regulatory network, the concentration of proteins is usually not measured due to high experimental cost. Therefore, the target of inference is to build a network consisting of measurable nodes. In regard to the gene regulatory network, this means the network is inferred on the transcriptional level.
	
	Assuming the first $p<n$ states are observable (i.e. $C=[I,\bold{0}]$), model~\eqref{ssm} is rewritten as follows:
\begin{equation}{\tiny
 \left[\begin{array}{c}y(t+1)\\h(t+1)\end{array}\right]=  \left[\begin{array}{cc}A_{11}&A_{12}\\A_{21}&A_{22}\end{array}\right] \left[\begin{array}{c}y(t)\\h(t)\end{array}\right]+ \left[\begin{array}{c}B_{u1}\\B_{u2}\end{array}\right]u(t)+ \left[\begin{array}{c}B_{e1}\\B_{e2}\end{array}\right]e(t),
}	
\end{equation}
where $h\in\mathcal{R}^{n-p}$ are hidden states. To avoid inferring hidden states, they are removed from the model. Dynamical structure functions (DSF) encode the information of hidden states via transfer functions~\cite{yuan}:
	\begin{equation}
	\begin{aligned}
	Y = QY + PU + HE.\\
	\end{aligned}
	\label{DSF}
	\end{equation}
	where $q$ denotes the time shift operator ($y(t+1)=qy(t)$) and:
	\begin{equation}
	\begin{aligned}
	Q &= (qI-D)^{-1}(W-D)\\
	P &= (qI-D)^{-1}V_u\\
	H &= (qI-D)^{-1}V_e. \\
	\end{aligned}
	\end{equation}
	with
	\begin{equation}
	\begin{aligned}
	W &= A_{11}+A_{12}(qI-A_{22})^{-1}A_{21}\\
	V_u &= A_{12}(qI-A_{22})^{-1}B_{u2}+B_{u1}\\
	V_e &= A_{12}(qI-A_{22})^{-1}B_{e2}+B_{e1}\\
	D &= diag\{W_{11},W_{22},...,W_{pp}\}.
	\end{aligned}
	\end{equation}	
$Q$, $P$ and $H$ are transfer matrices, each element of which is a strictly proper transfer function, indicating that the network is a causal system~\cite{yuan1}. Matrix $Q$ implies the connectivity among observable nodes, whose diagonal elements are zero. $P$ and $H$ matrices relate inputs and process noise to nodes, respectively. The topology of the network (i.e. model structure) is reflected by the zero structure of these three matrices. For example, if $[Q]_{ij}$ is zero, the $j$th node does not control the $i$th node. Model structures are denoted by $\mathcal{M}_k$ and $M_k$ presents the number of links. In particular, $\mathcal{M}_0$ represents the fully-connected topology. The internal dynamics of the network are described by the transfer functions. The order of a transfer function is relevant to the number of hidden states involved in that regulation pathway.
	
	The input-output map of the network is associated with the DSF as follows:
	\begin{equation}
	\begin{aligned}
	Y = G_uU + G_eE.
	\end{aligned}
	\end{equation}  
	where
	\begin{equation}
	\begin{aligned}
	G_u&=(I-Q)^{-1}P\\
	G_e&=(I-Q)^{-1}H.
	\end{aligned}
	\end{equation}
	Ideally, the input-output map can be perfectly recovered based on measurement data. Nevertheless, the corresponding DSF may not be unique, meaning that the network topology is unidentifiable. To ensure the inference problem is well-posed, additional constraints are imposed.
	\begin{prop}(Identifiability of DSF networks)~\cite{qpmodel}
		
		Given a $p\times (m+q)$ transfer matrix $G=[G_u, G_e]$, the DSF is identifiable if and only if $p-1$ elements in each column of $[Q, P, H]'$ are known, which uniquely specifies the component of $(Q,P,H)$ in the null space of $[G', I]$.
	\end{prop}
	
	A sufficient condition for network identifiability is that matrix $H$ is diagonal so that $p-1$ elements in each column of $[Q, P, H]'$ are known to be zero. In what follows, we make following assumptions so that no prior knowledge of matrix $P$ (structure of input channels) is required.
	\begin{assum}
		Noise matrix $H$ is diagonal, monic ($\lim_{q\rightarrow \infty} qH=I$) and minimal phase.
	\end{assum}
	
	Stability and sparsity are the basic nature of many practical networks such as biological networks and power systems. Therefore, we assume the target network is stable and sparse.
	\begin{assum}
		Each elements of transfer matrices, $Q$ and $P$ are stable. Matrices $Q$ and $P$ are sparse.
	\end{assum}
	
	\subsection{The likelihood distribution}
	After simple manipulations, the DSF in~\eqref{DSF} can be rewritten as:
\begin{equation}
	Y= F_yY +F_uU + \bar{E}.
	\label{pred}
	\end{equation}
	where
	\begin{equation}
	F_u = (qH)^{-1}P,~
	F_y =  I-(qH)^{-1}(I-Q),~
	\bar{E} = q^{-1}E.
	\end{equation}
	Model~\eqref{pred} is a valid causal system. According to the assumptions, transfer matrices, $F_u$ and $F_y$ are stable. In addition, since $H$ is diagonal, $F_u$ and $F_y$ have the same zero structure as $P$ and $Q$. As a result, $F_u$ and $F_y$ are sparse matrices and imply the network topology.
	
		Identifying model~\eqref{pred} is non-trivial. Since the number of hidden states and the connectivity among them are unknown, estimating the order of transfer functions requires an exhaustive search of all possible combinations, which is computationally prohibitive for large-scale networks. Additionally, imposing stable transfer matrices is problematic. To simplify the identification problem, we express model~\eqref{pred} in a non-parametric way. By doing so, the selection of model complexity is avoided and, more importantly, system stability can be promoted effectively. The dynamical system for the $i$th target node, is formulated below:
	\begin{equation}
	\begin{aligned}
	y_i(t) &= \sum_{j=1}^p\sum_{k=1}^{\infty}h_{ij}^y(k)y_j(t-k)\\
	&+ \sum_{j=1}^m\sum_{k=1}^{\infty}h_{ij}^u(k)u_j(t-k)+\bar{e}_i(t).
	\end{aligned}
	\label{nonpred}
	\end{equation}
	where $h_{ij}^y$ and $h_{ij}^u$ are the impulse responses of transfer functions $[F_y]_{ij}$ and $[F_u]_{ij}$, respectively. The objective is to estimate the impulse responses.
	
	Due to the implementation purpose, the impulse responses are truncated after sample time $T$. $T$ is set sufficiently large in order to catch the major dynamics of the impulse responses (i.e. $|h(k)|\approx 0$ for $k\geq T$).
	Assume the availability of time-series data collected from $L$ independent experiments for each node and input. For the $i$th target node with $\mathcal{M}_0$, we define following matrices and vectors. For other possible model structures, $\mathcal{M}_k$, the corresponding terms are defined in the same way.
	\begin{equation} {\tiny
	\begin{aligned}
	&Y_q = \left[\begin{array}{c}y_{q,i}(N_q)\\\vdots \\y_{q,i}(T_q+1)\end{array}\right], w_q = \left[\begin{array}{c}w_{q,1}\\ \hline \vdots \\ \hline w_{q,p+m}\end{array}\right]\\
	& \Phi_q = \left[\begin{array}{cc}\Phi_{q,y}&\Phi_{q,u}\end{array}\right]\\
	&\Phi_{q,y} =\left[\begin{array}{ccc} y_{q,1}(N_q-1:N_q-T_q)&\cdots&y_{q,p}(N_q-1:N_q-T_q)\\ \vdots &\ddots&\vdots \\y_{q,1}(T_q:1)&\cdots&y_{q,p}(T_q:1)\end{array}\right]\\
	&\Phi_{q,u} =\left[\begin{array}{ccc} u_{q,1}(N_q-1:N_q-T_q)&\cdots&u_{q,m}(N_q-1:N_q-T_q)\\ \vdots &\ddots&\vdots \\u_{q,1}(T_q:1)&\cdots&u_{q,m}(T_q:1)\end{array}\right]\\
	& \sigma^{-1} = E\{\bar{e}_i(t)^2\}.\\
	\end{aligned}}
	\label{para}
	\end{equation}
	where subscript $q$ is the index of experiments. Under different experimental conditions, data are produced from different internal dynamics (i.e. independent impulse responses) whilst the network topology is unchanged. $N_q$ is the number of data points. $Y_q\in R^{N_q-T_q}$ are time-series of the $i$th node. $w_q\in R^{T_q(p+m)}$ contain $p+m$ groups of impulse responses, each of which corresponds to a transfer function of $F_y$ or $F_u$. $\Phi_q \in R^{(N_q-T_q)\times T_q(p+m)}$ include time-series of all the nodes and inputs. $\sigma$ is the noise precision. Note that the dimension of these quantities varies with respect to the model structure.
	
	Based on Bayes' rules, the likelihood distribution of the $i$th target node with $\mathcal{M}_k$  is:
	\begin{equation}
	\begin{aligned}
	&p(Y\big|w,\sigma,\mathcal{M}_k)\\
	&=\prod_{q=1}^L(2\pi\sigma^{-1})^{-\frac{N_q-T_q}{2}}\exp\{-\frac{\sigma}{2}\|Y_q-\Phi_q w_q\|_2^2\}.
	\end{aligned}
	\label{like}
	\end{equation}
	
	\subsection{The prior distributions}
	
	Full Bayesian treatment deploys prior distributions for each random quantity to build up a hierarchical structure. The prior distributions play a similar role of penalties in regularized optimization problems. They are the key elements to incorporate prior knowledge and impose desired constraints.
	
	Since the impulse responses of model~\eqref{pred} are  stable (i.e. $\sum_{k=1}^{\infty}|h(k)|<\infty$), regularizations for stability are imposed to incorporate the prior knowledge. Kernel machines provide an effective way to construct a functional space as the feasible domain of stable impulse responses~\cite{foun}. A reproducing kernel Hilbert space (RKHS) is established using a proper kernel function, which contains stable impulse responses~\cite{kern}. The impulse responses of the model are estimated by solving a regularized optimization problem in that RKHS~\cite{nonp2}. 

	
	From the Bayesian perspectives, kernel machines can be formulated by introducing Gaussian processes for impulse responses and solving a maximum a posteriori problem (MAP)~\cite{gauss, nonp4, nonp5}. Therefore, we assume the impulse responses of the model are independent Gaussian processes whose covariance functions are Tuned/Correlated kernels (TC kernel). TC kernel has been widely used to characterize stable impulse responses~\cite{nonp8}. Other valid kernels include Diagonal/Correlated kernel (DC kernel)\footnote{ $k_{DC}(t,s;\beta_1,\beta_2)=\beta_1^{\frac{(t+s)}{2}}\beta_2^{|t-s|}$, $\beta_1\in(0,1)$ and $\beta_2\in(-1,1)$} and second order stable spline kernel (SS kernel)\footnote{ $k_{SS}(s,t;\beta)=\frac{\beta^{t+s+max(t,s)}}{2}-\frac{\beta^{3max(t,s)}}{6}$, $\beta\in(0,1)$}~\cite{nonp1, kern}. The reason why TC kernel is applied in this paper will be explained in the following sections. As a result, the prior distribution for $w$ is:
	\begin{equation}
	\begin{aligned}
	p(w|\lambda,\beta,\sigma,\mathcal{M}_k) = \prod_{q=1}^L\prod_{i=1}^{M_k}  \mathcal{N}(w_{q,i}|0,\sigma^{-1}\lambda_i^{-1}K_{q,i}).\\
	\end{aligned}
	\end{equation} 
	where $\beta$ are hyperparameters of TC kernels, which control the exponential decaying rate of impulse responses. $\lambda$ are scale variables of the kernel functions. In kernel machines, they introduces the effect of Automatic Relevance Determination (ARD) that promotes sparsity estimation~\cite{nonp}. In the Bayesian model, $\lambda$ influence the probability of model structure (i.e. network topology). As $\lambda_i$ approaches infinity, distribution $p(w_i|\lambda_i,\beta_i,\sigma)$ deploys an impulse at the origin, enforcing zero impulse responses. In this case, the $i$th node or input does not control the target node. Note that hyperparameters $\beta$ and $\lambda$ are shared in all experiments. As the standard setting of variational inference, noise precision $\sigma$ is also used to scale the covariance matrix. $K_{q,i}\in\mathcal{R}^{T_q\times T_q}$, $\lambda=[\lambda_1,...,\lambda_{M_k}]'$, $\beta=[\beta_1,...,\beta_{M_k}]'$ and 
	\begin{equation}
	\begin{aligned}
	\left[K_{q,i}\right]_{ts} &= k(t,s;\beta_i), ~k(t,s;\beta_i)= \beta_i^{max(t,s)}\\
	0< &\beta_i<1,~\lambda_i\geq0.
	\end{aligned}
	\end{equation} 
	Since $\sigma$ is non-negative, the Gamma distribution is assigned as its conjugate prior. Without specific preference on $\sigma$, parameters $a_0$ and $b_0$ of the distribution are set to $0.001$, resulting in a non-informative prior.
	\begin{equation}
	\begin{aligned}
	p(\sigma|a_0,b_0) = Gamma(\sigma|a_0,b_0)=\frac{b_0^{a_0}}{\Gamma(a_0)}\sigma^{a_0-1}e^{-b_0\sigma}.
	\end{aligned}
	\end{equation} 
	where $\Gamma(\cdot)$ is the gamma function.
	
	Finally, hyperpriors are assigned to hyperparameters to complete the hierarchy. For hyperparameter $\lambda_i$, the Gamma distribution is applied as the conjugate prior. Similar to $\sigma$, a non-informative prior is adopted.
	\begin{equation}
	\begin{aligned}
	p(\lambda_i|a_0,b_0) &= Gamma(\lambda_i|a_0,b_0)=\frac{b_0^{a_0}}{\Gamma(a_0)}\lambda_i^{a_0-1}e^{-b_0\lambda_i}.
	\end{aligned}
	\end{equation} 
For hyperparameter $\beta_i$, the uniform distribution on $(0,1)$ is employed as the prior, i.e., $	p(\beta_i) = 1,~ 0<\beta_i<1.$

	\subsection{The full Bayesian model}
By incorporating the likelihood and prior distributions, the full Bayesian distribution for model~\eqref{nonpred} is:
	\begin{equation}
	\begin{aligned}
	&p(w,\sigma,\lambda,\beta|Y,\mathcal{M}_k)\\ 
	&\propto\prod_{q=1}^L \left\{(2\pi \sigma^{-1} )^{-\frac{N_q-T_q}{2}}\exp\{-\frac{\sigma}{2}\parallel Y_q-\Phi_q w_q\parallel_2^2\}\right.\\
	&\left.\times|2\pi\sigma^{-1}\Lambda_q^{-1}K_q|^{-\frac{1}{2}}\exp\{ -\frac{\sigma}{2}w_q'\Lambda_q K_q^{-1}w_q\}\right\}\\
	&\times\frac{b_0^{a_0}}{\Gamma(a_0)}\sigma^{a_0-1}e^{-b_0\sigma}\times\prod_{i=1}^{M_k} \frac{b_0^{a_0}}{\Gamma(a_0)}\lambda_i^{a_0-1}e^{-b_0\lambda_i}.\\
	\end{aligned}
	\label{fullB}
	\end{equation} 
	where
$K_q=blkdiag\{K_{q,1},...,K_{q,M_k}\}$ and $\Lambda_q= diag\{\lambda\}\otimes I_{T_q}.$
	
	\subsection{Estimation of hyperparameters}
	
	Given~\eqref{fullB}, impulse responses are estimated as the mean of the marginal posterior distribution (i.e. $p(w|Y)=\int p(w,\sigma,\lambda,\beta|Y)d\sigma d\lambda d\beta$). Nevertheless, distribution $p(w|Y)$ is intractable because the full Bayesian model is highly nonlinear with respect to hyperparameters. Therefore, the estimate of impulse responses cannot be calculated in a closed form.
	
	In the system identification community, deterministic Bayesian approximations have been widely used as the remedy to accommodate non-gaussianity~\cite{pattern}. A candidate distribution, $p(w)$ is proposed to approximate the marginal distribution, $p(w|Y)$ analytically.  A metric is designed to measure the error between the approximate and the true distributions. The approximate distribution is optimized by minimizing the metric. Finally, the estimate of $w$ is calculated as the mean of the optimal candidate distribution.
	
	Empirical Bayes and variational inference are two typical methods of deterministic Bayesian approximations whilst empirical Bayes is more prevalent in the kernel-based system identification. The conditional distribution, $p(w|\lambda, \beta, \sigma,Y)$ is used to approximate the marginal distribution, $p(w|Y)$. The hyperparameters are optimized by solving a type II  or evidence maximization problem (i.e. $(\lambda_{opt}, \beta_{opt}, \sigma_{opt})=argmin_{\lambda, \beta, \sigma} -\log p(\lambda, \beta, \sigma|Y)$). Consequently, the optimal conditional distribution is $p(w|\sigma_{opt},\lambda_{opt},\beta_{opt},Y)$. The estimate of $w$ can be easily calculated as $\hat{w}=E_{w|\sigma,\lambda,\beta,Y}(w)$. 
	
 Empirical Bayes provides an effective way to estimate the hyperpameters of kernels. This framework has been shown very powerful in exploring system dynamics~\cite{nonp6}. Due to the effect of ARD, the estimated $\lambda$ can also be used for model selection (detection of network topology), leading to sparse networks.
	The zero structure of $\lambda^{-1}$ indicates the network topology. Nevertheless, empirical Bayes does not evaluate model evidence, $p(Y|\mathcal{M}_k)$. Hence, it is difficult to assess the relative confidence of the estimated model structure over the other possible structures. 
	
	Compared with empirical Bayes, VI provides a reasonable approximation of model evidence that is essential for topology detection.VI has been applied to estimate models that can be cast as a sparse  linear regression, where it has been shown that VI outperforms empirical Bayes via Monte Carlo simulations~\cite{VI1} (i.e. sparse Bayesian learning~\cite{prior1, tipp, jin, pan}). Nevertheless, VI is much less popular in kernel-based system identification. The updated factors in each iteration no longer have closed-form expressions due to the complex structure of kernel functions, which requests inner sampling loops. In addition, operations of high-dimensional matrices are more involved in this case. In particular, VI have to calculate the inversion and determinant of the ill-conditioned covariance matrix constructed from kernel functions. Nevertheless, by using TC kernel, the computational efficiency and robustness of VI are dramatically improved, which makes VI applicable to practical applications.
	
	\section{VARIATIONAL INFERENCE OF DYNAMICAL STRUCTURE FUNCTIONS}\label{sec:inference}
	
	\subsection{Update of random quantities}
	
	For each model structure, $\mathcal{M}_k$ of~\eqref{nonpred}, the corresponding full Bayesian model, $p(w,\sigma,\lambda,\beta|Y,\mathcal{M}_k)$ is approximated by a candidate distribution, $Q(w,\sigma,\lambda,\beta|\mathcal{M}_k)$ using the mean field factorization:
	\begin{equation}
	\begin{aligned}
	Q(w,\sigma,\lambda,\beta|\mathcal{M}_k) = q(w,\sigma|\mathcal{M}_k)q(\lambda|\mathcal{M}_k)q(\beta|\mathcal{M}_k).
	\end{aligned}
	\label{fact}
	\end{equation}   
	where $Q(\cdot)$ and $q(\cdot)$ are valid probability distributions. Hereafter, the symbol, $\mathcal{M}_k$ is suppressed to simplify the notation.
	
	The factors of~\eqref{fact}  are formulated according to~\eqref{q}. In what follows, the terms independent on the random variables of the factor under consideration are ignored for convenience. To begin with, factor $q(w,\sigma)$ is solved as the Gaussian-Gamma distribution:
	\begin{equation}
	\begin{aligned}
	&\ln q(w_q,\sigma)=\ln \mathcal{N}(w_q|\mu_q,\sigma^{-1}\Sigma_q)-\ln Gamma(\sigma|a^{\sigma},b^{\sigma}).
	\end{aligned}
	\end{equation}   	
	where
	\begin{equation}
	\begin{aligned}
	&\Sigma_q^{-1} = \Phi_q'\Phi_q+E_{\lambda}(\Lambda_q)E_{\beta}(K_q^{-1}),~\mu_q = \Sigma_q \Phi_q'Y_q\\
	&a^{\sigma} = \frac{\sum_{q=1}^L N_q-T_q}{2}+a_0,\\ &b^{\sigma} = b_0+ \frac{1}{2}\sum_{q=1}^L(Y_q'Y_q-\mu_q'\Sigma_q^{-1}\mu_q),\\
	&E_{\sigma,w}(\sigma w_qw_q') = \frac{a^{\sigma}}{b^{\sigma}}\mu_q \mu_q'+\Sigma_q,~
	E_{\sigma,w}(\sigma w_q) =  \frac{a^{\sigma}}{b^{\sigma}}\mu_q.
	\end{aligned}
	\end{equation}
	Following the same procedure, factor $q(\lambda)$ is solved as independent Gamma distributions:
	\begin{equation}
	\begin{aligned}
	&\ln q(\lambda_i) =  \ln Gamma(\lambda_i|a_{\lambda_i},b_{\lambda_i}).
	\end{aligned}
	\end{equation}
	where
	\begin{equation}
	\begin{aligned}
	&a_{\lambda_i}=\frac{\sum_{q=1}^LT_q}{2}+a_0,E_{\lambda}(\lambda_i) = \frac{ a_{\lambda_i}}{ b_{\lambda_i}}\\
	 &b_{\lambda_i}=b_0+\frac{1}{2}trace\left[\sum_{q=1}^L E_{\beta}(K_{q,i}^{-1})\overline{E_{w,\sigma}(\sigma w_qw_q')_i^{T_q}}\right].\\
	\end{aligned}
	\end{equation}
	Finally, factor $q(\beta)$ is calculated as:
	\begin{equation}
	\begin{aligned}
	\ln q(\beta)
	&=E_{w,\sigma,\lambda}[ \ln p(w|\sigma,\lambda,\beta)+\ln p(\beta) ]+ c_{w,\sigma,\lambda}\\
	&=E_{w,\sigma,\lambda}[ \ln p(w|\sigma,\lambda,\beta)]+\ln p(\beta) + c_{w,\sigma,\lambda}.\\
	\end{aligned}
	\end{equation}
	where
	\begin{equation}
	\begin{aligned}
	&E_{w,\sigma,\lambda}[ \ln p(w|\sigma,\lambda,\beta)]\\
	&=\sum_{q=1}^L- \frac{1}{2}\ln |K_q|-\frac{1}{2}trace\left[E_{\lambda}(\Lambda_q) K_q^{-1}E_{w,\sigma}(\sigma w_qw_q')\right],\\
	&\ln p(\beta) = 0.\\
	\end{aligned}
	\end{equation}
	Unlike the other random variables, factor $q(\beta)$ has no closed-form expression. Nevertheless, the elements of $\beta$ are independently distributed as:
	\begin{equation}{\tiny
	\begin{aligned}
	&q(\beta_i)\\
	&= c_i\prod_{q=1}^L|K_{q,i}|^{- \frac{1}{2}}\exp\left\{- \frac{1}{2}E_{\lambda_i}(\lambda_i)trace\left[ K_{q,i}^{-1}\overline{E_{w,\sigma}(\sigma w_qw_q')_i^{T_q}}\right]\right\}.\\
	\end{aligned}}
	\label{q_b}
	\end{equation}
	where $c_i$ is the unknown normalization constant.
	
	\subsection{Lower bound of model evidence}
	Combining all the factors above, the lower bound, $L[Q(w,\sigma,\lambda,\beta|\mathcal{M}_k)]$ of model evidence $p(Y|\mathcal{M}_k)$ is:
	\begin{equation}
	\begin{aligned}
	&L[Q(w,\sigma,\lambda,\beta|\mathcal{M}_k)]\\
	&=\frac{1}{2}\sum_{q=1}^L\ln |\Sigma_q| -\left[\frac{a_{\sigma}}{b_{\sigma}}\|Y_q-\Phi_q\mu_q\|_2^2+trace(\Phi_q\Sigma_q \Phi_q')\right]\\
	&-b_0\frac{a_{\sigma}}{b_{\sigma}}-a_{\sigma}\ln b_{\sigma}-\sum_{i=1}^{M_k} a_{\lambda_i}\ln b_{\lambda_i}+\frac{M_k\sum_{q=1}^LT_q}{2}\\
	&+M_k[a_0\ln b_0 -\ln \Gamma(a_0)]\\
	&-\sum_{i=1}^{M_k} b_0\frac{a_{\lambda_i}}{b_{\lambda_i}}-a_{\lambda_i}-\ln \Gamma(a_{\lambda_i})+\ln c_i.\\
	\end{aligned}
	\label{bound}
	\end{equation}
	where the constant terms independent on $\mathcal{M}_k$ are ignored. 
	
	\subsection{Algorithm for variational inference}

	Unfortunately, factors $q(w,\sigma)$, $q(\lambda)$ and $q(\beta)$ cannot be solved analytically. Hence, they are calculated cyclically in each iteration of the algorithm. The procedure is summarized in Algorithm~\ref{alg1}.
	\begin{algorithm}[!]
		\caption{Variational inference of DSF}
		\label{alg1}
		\begin{algorithmic}[1]
			\State Initialize $\mu$, $\Sigma$, $a_{\sigma}$, $b_{\sigma}$, $a_{\lambda_i}$,$b_{\lambda_i}$ and $E_{\beta}(K^{-1})$
			\For  {$k=1:Max$}
			\State Update $q(w,\sigma)$ as a Gaussian-Gamma distribution:
			\begin{equation}{\tiny
			\begin{aligned}
			\Sigma_q^{-1} &= \Phi_q'\Phi_q+E_{\lambda}(\Lambda_q)E_{\beta}(K_q^{-1})\\
			\mu_q &= \Sigma_q \Phi_q'Y_q\\
			a^{\sigma} &= \frac{\sum_{q-1}^LN_q-T_q}{2}+a_0\\
			b^{\sigma}&= b_0+ \frac{1}{2}\sum_{q=1}^L(Y_q'Y_q-\mu_q'\Sigma_q^{-1}\mu_q)\\
			\end{aligned}}
			\end{equation}
			\State Update $q(\lambda)$ as a Gamma distribution:
			\begin{equation}{\tiny
			\begin{aligned}
			a_{\lambda_i}&=\frac{\sum_{q=1}^LT_q}{2}+a_0\\
			b_{\lambda_i}&=b_0+\frac{1}{2}trace\left[\sum_{q=1}^LE_{\beta_i}(K_{q,i}^{-1})\overline{(\frac{a_{\sigma}}{b_{\sigma}}\mu_q\mu_q'+\Sigma_q)_i^{T_q}}\right]\\
			E(\lambda_i)&=\frac{a_{\lambda_i}}{b_{\lambda_i}}\\
			\end{aligned}}
			\end{equation}
			\State Update $q(\beta_i)$ and $E_{\beta_i}(K_i^{-1})$ according to:
			\begin{equation}{\tiny
			\begin{aligned}
			&q(\beta_i)\\
			&=c_i\prod_{q=1}^L|K_{q,i}|^{- \frac{1}{2}}\exp\left\{- \frac{a_{\lambda_i}}{2b_{\lambda_i}}trace\left[ K_{q,i}^{-1}\overline{(\frac{a_{\sigma}}{b_{\sigma}}\mu_q\mu_q'+\Sigma_q)_i^{T_q}}\right]\right\}\\
			\end{aligned}}
			\end{equation}
			\State Update the lower bound, $L[Q]$ of model evidence according to~\eqref{bound}.
			\If {$L[Q]_k-L[Q]_{k-1}<\epsilon$}
			\State Break
			\EndIf
			\EndFor
			\State Estimate $w$ as $\hat{w}=\mu$ and store $L[Q]$ for topology detection
		\end{algorithmic}
	\end{algorithm}
	
	In Algorithm~\ref{alg1}, term $E(K^{-1})$ is estimated in each iteration. However, since $q(\beta)$ is only known up to a normalization constant, $E(K^{-1})$ cannot be calculated analytically. Hence, numerical sampling methods are used to estimate the expectation. Since hyperparameters $\beta_i\in(0,1)$ are independently distributed, they can be sampled in parallel. We use the Metropolis-Hastings sampling method to draw samples from \eqref{q_b}. A uniform distribution is applied as the proposal distribution for $\beta_i$. Let $\hat{p}(\beta_i)=\prod_{q=1}^L|K_{q,i}|^{-\frac{1}{2}}\exp\{-\frac{a_{\lambda_i}}{2b_{\lambda_i}}trace[K_{q,i}^{-1}\overline{(\frac{a_{\sigma}}{b_{\sigma}}\mu_q\mu_q'+\Sigma_q)_i^{T_q}}]\}$. At current state $\beta_i^k$, a proposal $\beta_i^p$  is drawn from:
	\begin{equation}
	\begin{aligned}
	q_p(\beta_i^{p}|\beta_i^{k})=\left\{\begin{array}{ll}U(\beta_i^t-\frac{\varepsilon}{2},\beta_i^t+\frac{\varepsilon}{2}) &\frac{\varepsilon}{2}<\beta_i^{k}< 1-\frac{\varepsilon}{2}\\ U(0,\varepsilon)&\frac{\varepsilon}{2}\geq \beta_i^{k}\\U(1-\varepsilon,1)&1-\frac{\varepsilon}{2}\leq \beta_i^{k}\end{array}\right. .\\
	\end{aligned}
	\end{equation}
	where $U(a,b)$ is the uniform distribution on $(a,b)$. $\varepsilon$ is the selection window for sampling, which is set to $0.1$. The acceptance ratio is $r(\beta_i^p|\beta_i^k)=\min\{1,\frac{\hat{p}(\beta_i^p)q_p(\beta_i^k|\beta_i^p)}{\hat{p}(\beta_i^k)q_p(\beta_i^p|\beta_i^k)}\}$. If the proposal is accepted, $\beta_i^{k+1}=\beta_i^p$. Otherwise, $\beta_i^{k+1}=\beta_i^k$. With $N$ samples $\{\beta_i^k|k=1, ..., N\}$, $E(K_{q,i}^{-1})$ is estimated as $E(K_{q,i}^{-1})=\frac{1}{N}\sum_{k=1}^N (K_{q,i}^{k})^{-1}$.
	
	In addition, in order to calculate the lower bound of model evidence, one has to estimate the normalization constant, $c_i$ of $q(\beta_i)$ in~\eqref{q_b}. Since $\beta_i$ is a scalar on $(0,1)$,  numeric integration methods (e.g. adaptive quadrature~\cite{int}) are sufficient to give an accurate estimation based on $c_i^{-1}=\int \hat{p}(\beta_i)d\beta_i$.

	\subsection{Algorithm implementation}
	
	The algorithm requires to calculate the inversion and determinant of the covariance matrix, $K_{q,i}$ in each iteration. These two operations are computationally heavy because $K_{q,i}$ is a full matrix, and its inversion and determinant must be calculated thousands of times in the sampling loop of $\beta_i$ per iteration. With standard methods, they both demand $O(T_q^3)$ work. More importantly, $K_i$ may be ill-conditioned if hyperparameter $\beta_i$ is close to $0$, causing numerical instability during implementation~\cite{nonp10}. Therefore, it is necessary to find a robust and efficient way to deal with matrix operations. 
	
	It has been shown that the covariance matrix constructed using TC or DC kernels can be decomposed analytically whilst that of SS kernel cannot~\cite{nonp7, nonp11}. As a result, SS kernel is not adopted in our VI framework. Rather, TC kernel that possesses the simplest structure is applied to improve the robustness and to reduce the computational cost of the algorithm. The inverse covariance matrix, $K^{-1}\in\mathcal{R}^{T\times T} $ constructed using TC kernel (i.e. $[K]_{ts}=k(t,s;\beta)$) can be decomposed as~\cite{nonp8}:
	\begin{equation}
	\begin{aligned}
	K^{-1}&= U'WU.\\
	\end{aligned}
	\label{dec}
	\end{equation}
	where $U$ is a upper triangular matrix and $W$ is a diagonal matrix as follows:
	\begin{equation}
	\begin{aligned}
	U &= \left[\begin{array}{cccc}1 & -1 & \cdots& 0\\    0  & \ddots &\ddots&\vdots \\\vdots&0&\ddots&-1\\ 0&\cdots&0 & 1\end{array}\right]\\
	W &= (1-\beta)^{-1}diag\{\beta^{-1},\beta^{-2},...,\beta^{-T+1},\frac{1-\beta}{\beta^T}\}.
	\end{aligned}
	\end{equation} 
	Consequently, the matrix inversion only demands $O(T)$ work considering the sparse structure of $U$ and $W$, and the matrix determinant ($|K| = \beta^{\frac{T(T+1)}{2}}(1-\beta)^{T-1}$) requires $O(1)$.
	With $N$ samples of $\beta$ ($\{\beta^k|k=1,..,N\}$), the expectation of the inverse matrix costs $O(NT)$ work:
	\begin{equation}
	\begin{aligned}
	E(K^{-1})&=\frac{1}{N}U'\left[\sum_{k=1}^N W^{k}\right]U.\\
	\end{aligned}
	\end{equation}  
	
	\subsection{Detection of network topology}
	
	Hyperparameter $\lambda^{-1}$ are ARD variables, whose zero structure  determines the network topology. Nevertheless, due to local optimal solutions and numerical errors, none of these estimated variables are exactly zero in practical implementation. To improve the inference accuracy, the backward selection method is used for model selection. 
	
	Model selection is based on the posterior distribution of model structures (i.e. $p(\mathcal{M}_k|y)\propto p(y|\mathcal{M}_k)p(\mathcal{M}_k)$). Assuming equal probability for each model structure, the distribution is proportional to model evidence, $p(\mathcal{M}_k|y)\propto p(y|\mathcal{M}_k)$. The by-product of VI provides a lower bound of model evidence, which can be used to determine the most probable model structure (network topology). Nevertheless, the complete evaluation of model evidence requires an exhaustive search of all possible model structures, which is computationally prohibitive for large-scale networks. To tackle this problem, we come up with a heuristic strategy that applies backward selection to reduce the computational burden.
	
	To begin with, VI is conducted to infer a fully-connected network (i.e. $\mathcal{M}_0$). The confidence of inferred links is measured by the norm of their impulse responses. These links are removed from the model one-by-one, each time producing a particular model structure. The VI algorithm is implemented repeatedly with the proposed model structures until the network becomes empty. The best model structure is determined according to the highest lower bound of model evidence. The above procedure is summarized in Algorithm~\ref{alg2}.
	
	\begin{algorithm}[!]
		\caption{Variational inference of network topology}
		\label{alg2}
		\begin{algorithmic}[1]
			\State Implement Algorithm 1 with model structure $\mathcal{M}_0$.
			\State Set the threshold:
			\State $R=asc\{\sum_q[\|w_{q,1}\|,...,\|w_{q,n+p}\|]\}$.
			\For {k=1:n+p-1}
			\State $I = \{i|\sum_q\|w_{q,i}\|\leq R_k\}$
			\State Remove the links in index set $I$, producing $\mathcal{M}_k$
			\State Run Algorithm 1 with $\mathcal{M}_k$
			\State Store the value of $L[Q(w,\sigma,\lambda,\beta|\mathcal{M}_k)]$
			\EndFor
			\State $\mathcal{M}_{opt}=argmax_{\mathcal{M}_k}L[Q(w,\sigma,\lambda,\beta|\mathcal{M}_k)]$
		\end{algorithmic}
	\end{algorithm}

	\subsection{Theoretical analysis of the algorithm}
	Since the lower bound depends on four variables $\Sigma$, $\mu$, $b_{\lambda}$ and $c$ (other variables are either constant or determined by these four), let $\theta=[\Sigma,\mu,b_{\lambda},c]$. Define $\{\theta^k \}$ as the iterates generated using the coordinate descent. It is known that convergence of the lower bound is guaranteed if all iterates are calculated explicitly without approximations~\cite{pattern,VI1}. This section shows that the proposed algorithm still converges even with stochastic approximations (MCMC) used in our framework. In addition, the proposed algorithm effectively imposes sparse topologies. In what follows,  $\{\theta^k \}$ denote iterates without approximations whereas $\{\theta^k_{N_1,\cdots,N_{k-1}} \}$ are produced from the proposed algorithm where $N_k$ is the number of samples generated by MCMC in the $k$th iteration.
	\begin{prop}
		Define following sequences $\{\Sigma_M \}$, $\{\mu_M \}$, $\{[b_{\lambda}]_M \}$, $\{c_M\}$,  $\{q_M(\beta;\Sigma_M, \mu_M, [b_{\lambda}]_M, c_M)\}$ and $q(\beta;\Sigma, \mu, b_{\lambda}, c)$. If sequences are such that    $\lim_{M\rightarrow\infty}\Sigma_M =\Sigma$, $\lim_{M\rightarrow\infty}\mu_M=\mu$ and $\lim_{M\rightarrow\infty}[b_{\lambda}]_M=b_{\lambda}$,  normalization constants $c$ and $\{c_M\}$ are well defined, and $\lim_{M\rightarrow\infty}c_M=c$.  Assuming Markov chains $\{\beta_M^k\}$ sampled from $q_M(\beta)$ are Harris recurrent, we have $\lim_{M,N\rightarrow\infty} \frac{1}{N}\sum_{k=1}^NV(\beta_M^k)=E_q[V(\beta)]$ with probability $1$ for any bounded continuous function $V(\cdot)$.
		\label{preprop}
	\end{prop}
		Proof:
		Without loss of generality, consider the distribution $q(\beta)$~\eqref{q_b} from a single experiment:
		\begin{equation}{
			\begin{aligned}
			&q(\beta)
			= c|K|^{- \frac{1}{2}}\exp\left\{- \frac{1}{2}E_{\lambda}(\lambda)trace\left[ K^{-1}E_{w,\sigma}(\sigma ww')\right]\right\}.\\
			\end{aligned}}
		\end{equation}
		By using the decomposition of matrix $K$~\eqref{dec}, the above expression can be expanded as:
		\begin{equation}
		\begin{aligned}
		&q(\beta;c, \alpha) = cp(\beta;\alpha)  ,\\
		&p(\beta;\alpha)
		= \prod_{j=1}^TW_{jj}^{ \frac{1}{2}}\exp\left\{- \frac{1}{2}\alpha_jW_{jj}\right\},\\
		&c^{-1} = \int_0^1 p(\beta)d\beta,\\
		&W_{jj} = \left\{ \begin{array}{cc}\frac{1}{\beta^j(1-\beta)}&j<T\\ \frac{1}{\beta^T}&j=T\end{array}\right..\\
		\end{aligned}
		\end{equation}
		where $\alpha>0$ is a function of $\Sigma$, $\mu$, and $b_{\lambda}$. According to $(28)$ and $(30)$, and under the conditions of the proposition, we have $\lim_{M\rightarrow\infty}\alpha_{M}=\alpha$.\newline
		Consider the following function $f(x)$ with $x\in [0,+\infty)$ and $\alpha>0$:
		\begin{equation}{
			\begin{aligned}
			f(x)&=x^{ \frac{1}{2}}\exp\left\{- \frac{1}{2}\alpha x\right\}.\\
			\end{aligned}}
		\end{equation}
		Clearly, $f(x)$ is non-negative and attains its maximum at $x^{\star}=\frac{1}{\alpha}$ with $f(x^{\star})=\frac{1}{\sqrt{e\alpha}}$. Hence, $p(\beta;\alpha)$ is also bounded. Since $p(\beta;\alpha)$ is continuous, it is Riemann integrable on $[0,1]$ and hence Lebesgue integrable. Since $\lim_{M\rightarrow\infty} \alpha_{M} = \alpha$, we have $\lim_{M\rightarrow\infty} p_M(\beta;\alpha_M)=p(\beta;\alpha)$ pointwise. As $\{\alpha_{M} \}$ converges, there exist $\underline{\alpha}$ and $m_{\underline{\alpha}}$ such that $\alpha_{M}\geq\underline{\alpha}>0$ for $M>m_{\underline{\alpha}}$. As a result, $p_M(\beta;\alpha_M)\leq p(\beta;\underline{\alpha})$ for $\beta\in[0,1], M>m_{\underline{\alpha}}$. Since $p(\beta;\underline{\alpha})$ is Lebesgue integrable, $\lim_{M\rightarrow\infty}\int_0^1 p_M(\beta;\alpha_M)d\beta=\int_0^1 p(\beta;\alpha)d\beta$ according to the dominated convergence theorem. Eventually, $c$ and $c_M$  are well defined, and $\lim_{M\rightarrow\infty} c_M=\lim c$. \newline		
		It is then clear that $\lim _{M\rightarrow\infty}q_M(\beta;c_M,\alpha_M)=q(\beta;c,\alpha)$ pointwise. According to the Scheffe's theorem, random variables $\beta_M$ (distributed as $q_M(\beta)$) converge to $\beta$ in distribution. Therefore, $\lim_{M\rightarrow\infty} E_{q_M}[V(\beta_M)]=E_{q}[V(\beta)]$. With Harris recurrent Markov chains, we have $\lim_{N\rightarrow\infty} \frac{1}{N}\sum_{k=1}^NV(\beta_M^k)=E_{q_M}[V(\beta_M)]$ with probability $1$. Assuming MCMC proposals are well designed so that the convergence is uniform in $M$, we have $\lim_{M,N\rightarrow\infty} \frac{1}{N}\sum_{k=1}^NV(\beta_M^k)=E_q[V(\beta)]$ based on the theory of double sequence. 		 \hfill $\blacksquare$\newline
		With Proposition~\ref{preprop}, we are able to prove the algorithm convergence.
		\begin{prop}
			Lower bound $L(\theta_{N_1,\cdots,N_{k-1}}^k)$ converges to its maximum $L^{\star}=\max L(\theta)$ with probability $1$ as the number of samples ($N_k$) and iterations ($k$) approaches infinity.
		\end{prop}
	Proof: Since $L[Q(\theta)]$ is convex with respect to each factor $q_i(\theta_i)$, convergence of the lower bound is guaranteed using the coordinated descent method if updates are calculated explicitly~\cite{pattern,block}. Hence, $\lim_{k\rightarrow\infty}L(\theta^k)=L^{\star}$ where $\theta^k_j=\arg\min L(\theta_j;\theta^{k}_{i< j},\theta^{k-1}_{i> j})$ (parameters are updated in sequence). However, in this paper, some components $\theta^k_j$ are intractable. Rather, they are estimated using sampling methods. The objective is to show that $\lim_{N_{1},\cdots, N_{k-1}\rightarrow\infty}\theta_{N_1,\cdots,N_{k-1}}^{k}=\theta^{k}$ for $\forall k$. The proof is conducted using mathematical induction. Assuming $\lim_{N_{1},\cdots, N_{k-2}\rightarrow\infty}\theta_{N_1,\cdots,N_{k-2}}^{k-1}=\theta^{k-1}$ holds at iteration $k-1$, $\lim_{N_{1}\cdots N_{k-1}\rightarrow\infty}\theta_{N_1,\cdots,N_{k-1}}^{k}=\theta^{k}$ also holds at iteration $k$ according to Proposition~\ref{preprop}. In addition, we have $\lim_{N_{1}\rightarrow\infty}\theta^2_{N_1}=\theta^2$  with probability 1 after initialization $\theta^1$. Hence, $\lim_{N_{1},\cdots, N_{k-1}\rightarrow\infty}\theta_{N_1,\cdots,N_{k-1}}^{k}=\theta^{k}$ for $\forall k$. Since lower bound~\eqref{bound} is continuous with respect to all parameters, we have 
	$\lim_{k\rightarrow\infty}\lim_{N_{1},\cdots, N_{k-1}\rightarrow\infty}L(\theta_{N_1,\cdots,N_{k-1}}^{k})=L^{\star}$.
\hfill $\blacksquare$\newline
    Finally, we argue that the proposed algorithm promotes sparse topologies and is at least competitive with empirical Bayes~\cite{block,nonp}.
    \begin{rem}
    	The proposed algorithm imposes sparse topologies and is at least as efficient as empirical Bayes.  For any fixed $a^{\sigma}$, $b^{\sigma}$ and $q(\beta)$, the proposed algorithm only needs to update $E(\lambda)$ which determines the sparsity of topologies since $w_i\rightarrow 0$ as $E^{-1}(\lambda_i)\rightarrow 0$. Consider the following optimization problem derived from empirical Bayes~\cite{block}:
   \begin{equation}
   \begin{aligned}
   &\arg\min_{\gamma} \hat{Y}'(\sigma I+\hat{\Phi}\Gamma \hat{K}\hat{\Phi}')^{-1}\hat{Y}+\ln |\sigma I+\hat{\Phi}\Gamma \hat{K}\hat{\Phi}'|.
   \end{aligned}
   \end{equation}
   where $\sigma =( \frac{a^{\sigma}}{b^{\sigma}})^{-\frac{1}{2}}$, $\hat{\Phi}=(\frac{a^{\sigma}}{b^{\sigma}})^{-\frac{1}{4}}\Phi$, $\hat{Y}=(\frac{a^{\sigma}}{b^{\sigma}})^{\frac{1}{4}}Y$ and $\hat{K} = E^{-1}({K^{-1}})$.  $\Gamma$ is a block diagonal matrix whose zero structure determines the sparsity of topologies.\newline
   By using the expectation maximization method (EM) to solve the problem, the resulting update of $\Gamma$ is exactly the same as $E^{-1}(\lambda)$ in the proposed algorithm. 
   \end{rem}

	\section{SIMULATION}\label{sec:Simulation}
	
	We conducted Monte Carlo simulations and compared our method with another kernel-based system identification approach. The state-of-the-art method applies empirical Bayes to estimate hyperparameters and to detect network topology, where DC (Kernel\_DC), SS (Kernel\_SS) and TC (Kernel\_TC) kernels were all used for inference. KEB solves the optimization problem as follows.
	More details can be found in~\cite{nonp}.
	\begin{equation}
	\begin{aligned}
	&\arg\min_{\sigma,\gamma,\beta} Y'(\sigma I+\Phi\Gamma K\Phi')^{-1}Y+\ln |\sigma I+\Phi\Gamma K\Phi'|.
	\end{aligned}
	\label{EB}
	\end{equation}
	where $\Gamma$ is equivalent to $\Lambda^{-1}$ in our framework. $K$ is the covariance matrix constructed by kernel functions.
	If $\gamma_i$ is $0$, the $i$th node does not control the target node. To select the inferred links, a similar backward selection method is used~\cite{nonp}.
	 
	 The DSF  networks for test were generated randomly with diverse topologies (including an extremely sparse type, a ring structure), simulated under various noise levels and inferred using time-series data of different lengths.

	Two criteria are applied to evaluate the performance of algorithms, namely True Positive Rate (TPR) and Precision (PREC). TPR shows the percentage of the true links in the ground truths that were successfully inferred. TPR implies the information richness of the inference result. PREC equates to the rate of the correct links over the all inferred. PREC indicates the reliability or accuracy of the inferred network. Hence, ensuring a high PREC is the first priority in real applications. If, for example, PREC is below $50\%$, one cannot tell which inferred links are correct.	To investigate whether the infer networks have internal dynamics consistent with the ground truths, the estimated models were simulated to predict the validation dataset that was not used for inference. The prediction accuracy is measured based on the metric as follows.
	\begin{equation}
	fitness = 100\left(1-\frac{\|y-\hat{y}\|}{\|y-\bar{y}\|}\right)
	\end{equation}
	where $y$ are the validation data of a certain node, $\hat{y}$ are the predicted output and $\bar{y}$ are the mean of the validation data. The fitness of all nodes was averaged to generate the conclusion.

	\subsection{Random DSF networks}
	
	$100$ networks were generated with random topologies and internal dynamics. All networks contained $15$ nodes in total. Each node was independently driven by an input that was known and process noise that was assumed to be unknown during inference. DSF models were produced from state space models~\eqref{ssm}. To be specific, a sparse stable matrix $A\in\mathcal{R}^{15\times15}$ was first yielded randomly using the function $sprandn(n,n,density)$ in Matlab. The brute force strategy was applied to guarantee that matrix $A$ was Hurwitz (that is, no eigenvalue was outside the unit circle of the complex plane) and that no isolated nodes existed in the network. 
	
\begin{figure}
		\centering
		\includegraphics[width=0.7\columnwidth]{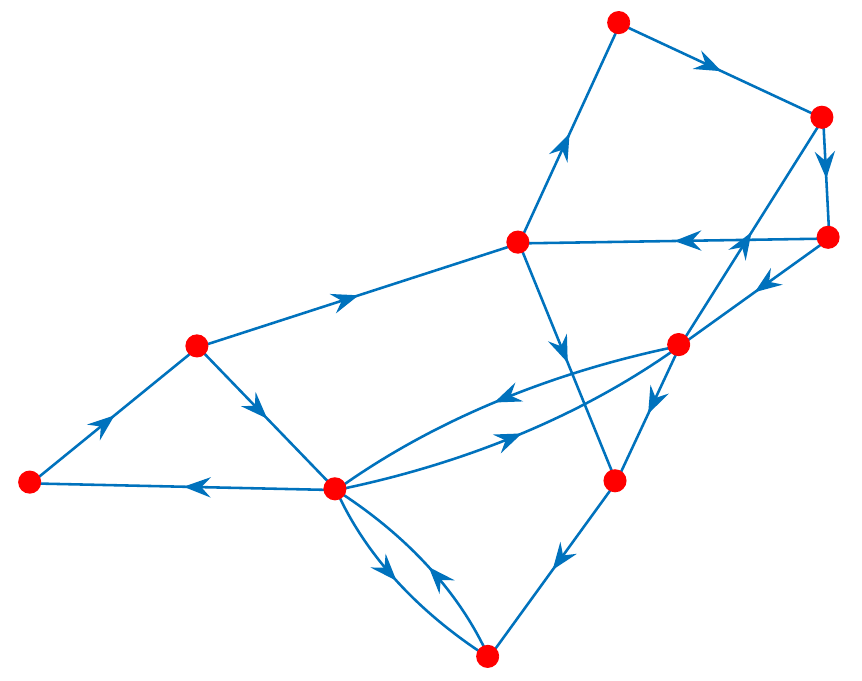}
		\caption{The structure of a randomly generated network. Solid lines with arrows represent links. Red circles denote nodes.}
		\label{random}
	\end{figure}

	To simulate the models, inputs and process noise were both i.i.d. white Gaussian signals. The variance of inputs was fixed to $1$ whilst that of process noise varied. The Signal-to-Noise ratio is defined as $SNR=10\log\frac{\sigma_u}{\sigma_e}$ where $\sigma_u$ and $\sigma_e$ are signal variance of inputs and noise, respectively. The first $10$ states of the models were measured, leaving the rest $5$ as hidden nodes. Figure~\ref{random} displays one example of the resulting networks. The length of  impulse responses after truncation was set to 20. The data for inference were collected with different lengths.
	
	The average TPR and PREC over $100$ trials are recorded in Table~\ref{M1}-\ref{M3}. In the best-case scenario (no noise), VI outperforms all the other methods. In particular, almost all the inferred links of VI are correct ($PREC\approx100\%$), regardless of the number of data points. Meanwhile, VI is able to identify most true links of the ground truths. 
	
	With $85$ data points, VI recovers the network perfectly. In contrast, Kernel\_TC presents the weakest result. PREC of Kernel\_TC stays low unless more data points are available for inference. The poor performance of Kernel\_TC indicates the effectiveness of VI that uses the same kernel function.
	
	As $SNR$ decreases to $10dB$, all methods require more data points to counteract the interference of process noise. Although TRP of VI is slightly lower than Kernel\_SS, PREC of VI is much higher than Kernel\_SS, which is close to $100\%$. Kernel\_TC also achieves accurate results. Nevertheless, many true links are missed ($TPR<80\%$).
	
	It is remarkable that the inferred networks of VI are highly reliable ($PREC\approx100\%$) even under the worst-case scenario (that of pure noise). The lack of data points only affects TPR of VI whilst PREC remains high. The gap of TPR between Kernel\_SS, Kernel\_DC and VI decreases gradually as more data points are available. Similar to the above cases, Kernel\_TC is outperformed by VI. 
	\begin{figure}
		\centering
		\includegraphics[width=1\columnwidth]{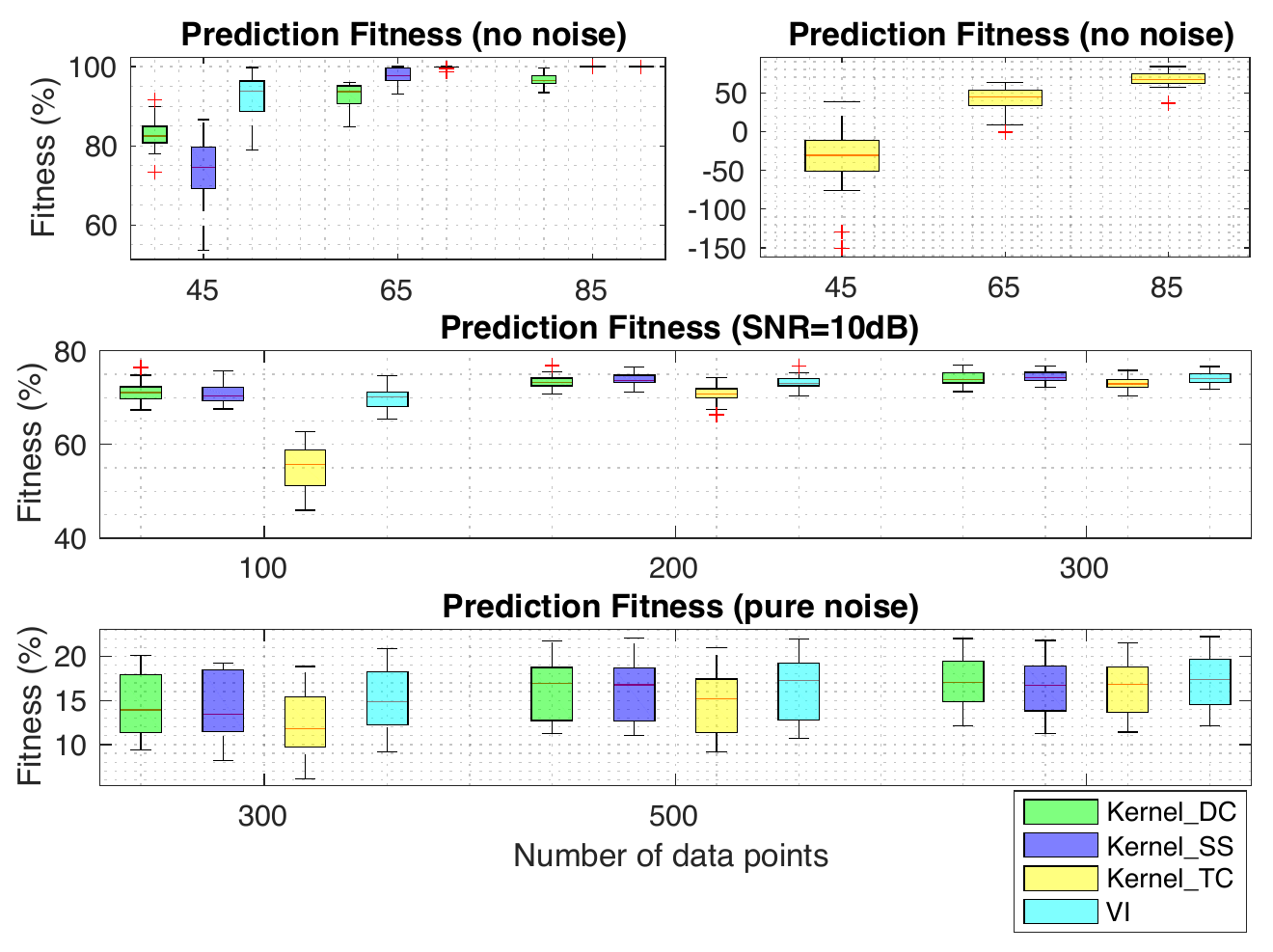}
		\caption{Prediction of randomly generated networks.}
		\label{fit_QP}
	\end{figure}
	Based on the simulations, VI presents great advantages on inference accuracy over the other methods. Almost all the inferred links of VI are correct regardless of the noise level and number of data points. Generally, TPR of VI is slightly lower than Kernel\_SS. However, VI actually missed only a few more true links compared with Kernel\_SS since the target networks were sparse. On average, there were $18.25$ links per network in the simulations. According to the results, VI, at most missed, $2.7$ links more than Kernel\_SS. 
	
The validation result is shown by the box plot in Figure~\ref{fit_QP}. With negligible process noise, VI clearly outperforms all the other methods. The prediction fitness of VI is above $80\%$ and reaches approximately $100\%$, given enough data. As $SNR$ decreases to $10dB$, all methods, except Kernel\_TC present similar performance. When process noise overwhelms inputs, the prediction fitness of all methods drops seriously to below $30\%$. In this case, the performance of VI is slightly better than the others. Note that Kernel\_TC presents the weakest result under different noise levels, implying that VI outperforms KEB at least with TC kernel.
	\begin{table}[h!]
		\caption{Inference of random networks with no noise}
		\centering
		\resizebox{\columnwidth}{!}{
			\begin{tabular}{ |c|c|c|c|c|c|c|}
				\hline
				\multicolumn{7}{|c|}{No Noise}\\
				\hline
				\multirow{2}{2em}   &\multicolumn{2}{|c|}{45} &\multicolumn{2}{|c|}{65}&\multicolumn{2}{|c|}{85} \\
				\cline{2-7}
				&PREC &TPR&PREC&TPR&PREC&TPR\\
				\hline
				Kernel\_DC      &  94.3	&56.9&97.7	&73.1&	97.9	&85.2\\
				Kernel\_SS      &  84.7	&58.3&	91.4&	91.7&	100	&100\\
				Kernel\_TC      &43.3	&16.5&	71.1&	23.4&	98.3&	42.9\\
				VI                   &100	&75.0&	99.7&	98.5&	100	&100\\
				\hline
			\end{tabular}
		}
		\label{M1}
	\end{table}
	
	
	\begin{table}[h!]
		\caption{Inference of random networks with $10dB$ SNR}
		\centering 
		\resizebox{\columnwidth}{!}{
			\begin{tabular}{ |c|c|c|c|c|c|c|  }
				\hline
				\multicolumn{7}{|c|}{10dB}\\
				\hline
				\multirow{2}{2em}   &\multicolumn{2}{|c|}{100} &\multicolumn{2}{|c|}{200}&\multicolumn{2}{|c|}{300} \\
				\cline{2-7}
				&PREC &TPR&PREC&TPR&PREC&TPR\\
				\hline
				Kernel\_DC      & 95.6	&77.1&	98.6&	84.2&	96.5&	86.9\\
				Kernel\_SS      & 74.1	&82.5&	79.7&	88.3&	86.3&	91.0\\
				Kernel\_TC      &92.5	&46.5&	100	&66.2	&99.6&	75.7\\
				VI                   &100&	68.2&	99.7	&81.3&	100	&86.4\\
				\hline
			\end{tabular}
		}
		\label{M2}
	\end{table}
	

	\begin{table}[h!]
		\caption{Inference of random networks with pure noise}
		\centering
		\resizebox{\columnwidth}{!}{
			\begin{tabular}{ |c|c|c|c|c|c|c|  }
				\hline
				\multicolumn{7}{|c|}{No Input}\\
				\hline
				\multirow{2}{2em}   &\multicolumn{2}{|c|}{300} &\multicolumn{2}{|c|}{500}&\multicolumn{2}{|c|}{1000} \\
				\cline{2-7}
				&PREC &TPR&PREC&TPR&PREC&TPR\\
				\hline
				Kernel\_DC      &77.6&72.9	&	89.1&	75.6&	96.9&	79.1\\
				Kernel\_SS      &64.3&69.6	&	81.7&	76.1&	88.0&	77.1\\
				Kernel\_TC      &77.9	&60.3	&87&65.7&93.4&70.9			\\
				VI                   &100&56.5&	100	&65.0&	100	&74.9\\
				\hline
			\end{tabular}
		}
		\label{M3}
	\end{table}
	
	\subsection{Ring networks}
	
\begin{figure}
		\centering
		\includegraphics[width=0.7\columnwidth]{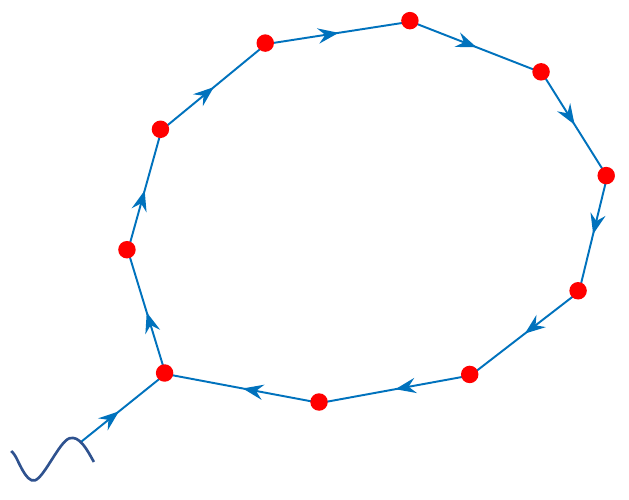}
		\caption{A network with the ring structure. Symbol '$\sim$' denotes the input signals.}
		\label{ring} 
	\end{figure}

	$100$ networks with the fixed ring structure as shown in Figure~\ref{ring} were generated and simulated following the same protocol in the last section. Each node was driven by independent process noise. Only one input entered the network through a single node. Since the network contains a feedback loop and is extremely sparse, it is more challenging to infer.
	
	Table~\ref{R2} presents the inference result. VI is still able to produce reliable networks with the highest PREC among all the methods ($PREC=100\%$). More importantly, PREC of the other three cases highly relies on the number of data points whilst that of VI does not. TPR of VI and Kernel\_SS is very close. Since the ring network contained only $10$ links, VI at most missed $3$ true links.
	
	Simulations indicate that VI is able to provide reliable inference results and identify most true links of the ground truths even if the target networks are extremely sparse. More importantly, the performance of VI is robust ($PREC\approx100\%$), which is crucial in real applications where the measurements are not sufficient for inference.
	
\begin{figure}
		\centering
		\includegraphics[width=1\columnwidth]{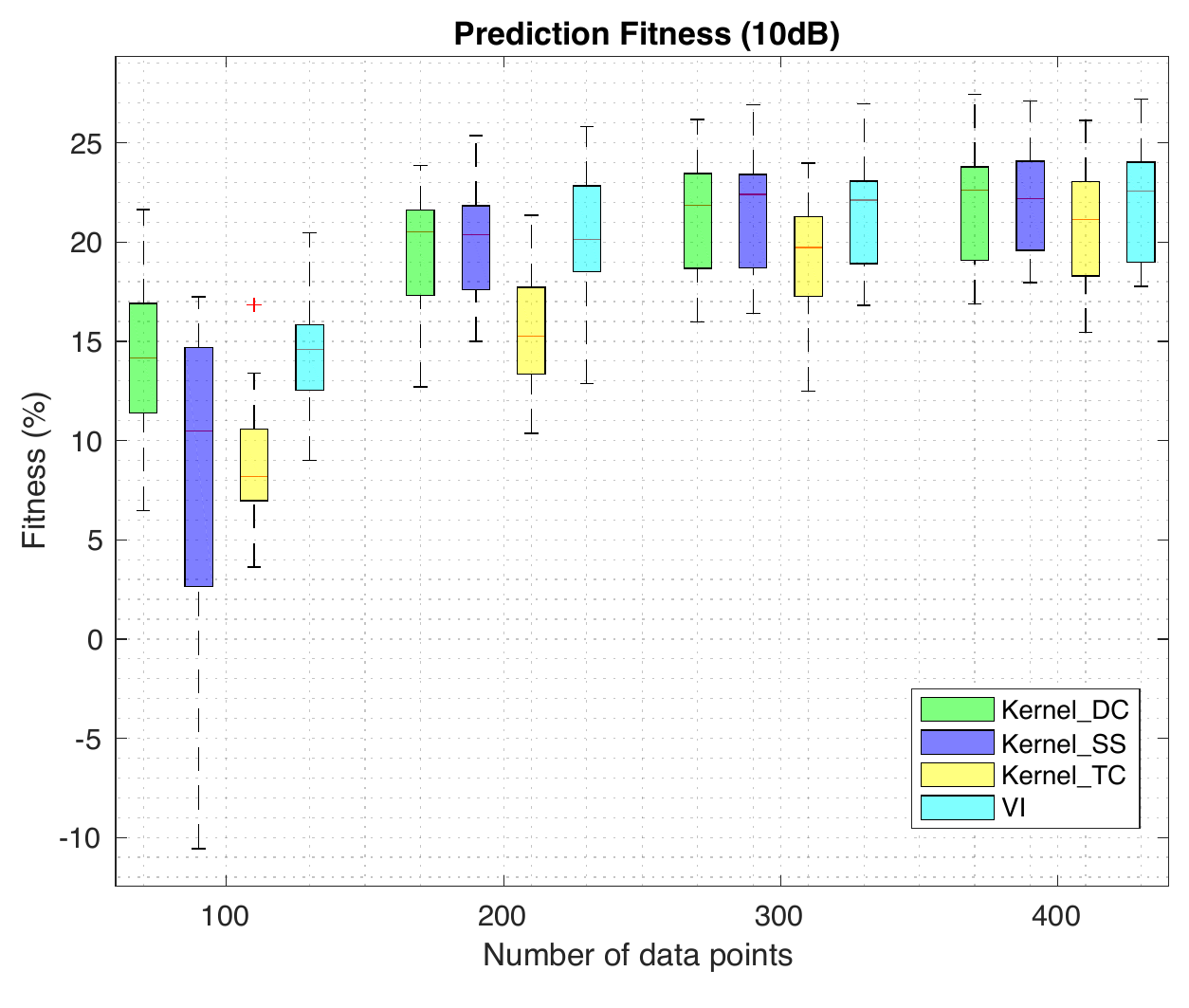}
		\caption{Prediction of ring networks.}
		\label{fit_Ring}
	\end{figure}

	The validation result is shown in~Figure \ref{fit_Ring}. Since the ring network was mostly driven by process noise, the prediction fitness of all methods is below $30\%$. The prediction accuracy of VI is competitive with Kernel\_DC. Kernel\_TC presents the weakest result.

\begin{table}[h!]
	\caption{Inference of ring networks with $10dB$ SNR}
	\centering
	\resizebox{\columnwidth}{!}{
		\begin{tabular}{ |c|c|c|c|c|c|c|c|c|  }
			\hline
			\multicolumn{9}{|c|}{10dB }\\
			\hline
			\multirow{2}{2em}   &\multicolumn{2}{|c|}{100} &\multicolumn{2}{|c|}{200}&\multicolumn{2}{|c|}{300} &\multicolumn{2}{|c|}{400}\\
			\cline{2-9}
			&PREC &TPR&PREC&TPR&PREC&TPR&PREC&TPR\\
			\hline
			Kernel\_DC      &51.0&75.0		&75.8&	81.0	&86.2	&84.0&93.8&84.5\\
			Kernel\_SS      &42.2&76.5	&64.2&	83.5	&67.8	&86.5&74.6&88.5\\
			Kernel\_TC      &76.9	&19.3	&99.3&32.5&97.3&	55.0&98.5&68.5		\\
			VI                   &100&	27.5&	100	&66.0	&100&	73.5&100&80.0\\
			\hline
		\end{tabular}
	}
	\label{R2}
\end{table}

	

	\section{CONCLUSION AND DISCUSSION}\label{sec:Conclusion}
	
	This paper has applied variational inference to identify DSF models given measured time series data. No prior knowledge of the hidden nodes including their number and internal connectivity is required. Both the system dynamics and topology of sparse linear networks can be inferred accurately. The proposed method achieves this by incorporating kernel-based methods to promote system stability and by introducing VI to imposing network sparsity. By decomposing the kernel function, the resulting algorithm becomes computationally efficient and robust. Monte Carlo simulations imply that our method always produces reliable inference result regardless of challenging experimental conditions (for example, lower number of data points, high levels of noise, and extremely sparse topologies). The inference of links is highly accurate (with nearly $100\%$ confidence): only a few true links are missed. Therefore, the developed method appears highly reliable for real-world applications.
	
	Overall, the value of this approach is that it outperforms  KEB at least in regard to TC kernel. Our method raises the reliability of inference results to the highest level ($100\% PREC$). In particular, out method is applicable to real-world networks where full state measurements are normally unavailable such as gene regulatory networks.  
	
	The performance of our method may be further improved using other kernel functions (e.g. DC and SS). However, the computational cost of doing so is heavy and the robustness of the algorithm is not guaranteed.
	
	Future developments should include two aspects. The first is to find an effective decomposition for other kernel functions so that they can be used in our framework. The second aspect is to further improve TPR of inferred networks while maintaining high PREC. Considering most real-word networks are nonlinear, it is necessary to extend our framework to black-box nonlinear systems.

	\bibliographystyle{plain}        
	\bibliography{autosam}           
	
%
%

\end{document}